# Frequency dependent superfluid stiffness in the pseudogap regime in strongly disordered NbN thin films

Mintu Mondal[a*], Anand Kamlapure[a], Somesh Chandra Ganguli[a], John Jesudasan[a], Vivas Bagwe[a], Lara Benfatto[b] and Pratap Raychaudhuri[a†]

[a]Tata Institute of Fundamental Research, Homi Bhabha Road, Colaba, Mumbai 400005, India.
[b]ISC-CNR and Department of Physics, Sapienza University of Rome, Piazzale Aldo Moro 5, 00185, Rome, Italy.

*Abstract:* We measure the frequency dependence of the complex ac conductivity of NbN films with different levels of disorder in frequency range 0.4-20 GHz. Films with low disorder exhibit a narrow dynamic fluctuation regime above $T_c$ as expected for a conventional superconductor. However, for strongly disordered samples, the fluctuation regime extends well above $T_c$, with a strongly frequency-dependent superfluid stiffness which disappears only at a temperature $T_m^*$ close to the pseudogap temperature obtained from scanning tunneling measurements. Such a finite-frequency response is associated to a marked slowing down of the superconducting fluctuations already below $T_m^*$. The corresponding large length-scale fluctuations suggest a scenario of thermal phase fluctuations between superconducting domains in a strongly disordered s-wave superconductor.

[*] mondal@tifr.res.in
[†] pratap@tifr.res.in

At the microscopic level, the remarkable properties of the superconducting state are characterized by two properties: (i) Pairs of electrons which form bound states, i.e. Cooper pairs, and (ii) the Cooper pairs which condense in a macroscopic phase coherent state. The former manifests as a gap in the electronic excitation spectrum, namely, the superconducting energy gap, $\Delta$, and the latter results in finite superfluid stiffness, $J$, which is the energy cost of twisting the phase of the condensate. Based on the Bardeen-Cooper-Shrieffer theory[1], which has been remarkably successful in explaining the properties of conventional superconductors, it has long been assumed that the superconducting transition temperature, $T_c$, is governed by $\Delta$ alone, whereas phase fluctuations only play a role infinitesimally close to $T_c$. Consequently, it has been justified to expect the superconducting transition to happen at the temperature where $\Delta$ goes to zero[2]. However, this central paradigm has recently been questioned in a series of experiments on strongly disordered $s$-wave superconductors[3,4,5] (TiN, InO$_x$ and NbN), all of which reveal the existence of a pseudogap (PG) state, where the gap in the electronic excitation spectrum continue to persist in the tunneling density of states (t-DOS) and disappears at a temperature $T^*$, well above $T_c$. Scanning tunneling spectroscopy (STS) measurements further reveal that at strong disorder the superconducting state become inhomogeneous, segregating into domains where the superconducting order-parameter is large and regions where the order-parameter is completely suppressed[5,6,7,8]. On the other hand disorder scattering is expected to reduce $J$, thereby rendering the system susceptible to phase fluctuations[9]. Therefore, these observations lead to the obvious question on whether strong disorder can destroy the superconducting state through phase fluctuations between domains, without significantly affecting the underlying pairing mechanism, thereby giving rise to a state with finite density of Cooper pairs but no global superconductivity.

It is therefore instructive to try to look remnants of superconductivity in the dynamic response of phase stiffness in the PG regime.

In this letter, we address this problem through measurement of the ac complex conductivity, $\sigma(\omega) = \sigma'(\omega) + i\sigma''(\omega)$, on NbN thin films with different levels of disorder using a broadband Corbino microwave spectrometer[10] in the frequency range 0.4 - 20 GHz. The advantage of this technique is that it is sensitive to temporal correlations and to the length scale set by the probing frequency. In the superconducting state where phase coherence extends over all length and time scales, the superfluid density ($n_s$) and hence $J$ can be determined from $\sigma''(\omega)$ using the relations[2,9],

$$\sigma''(\omega) = \frac{n_s e^2}{m\omega} \text{ and } J = \frac{\hbar^2 n_s a}{4m}, \qquad (1)$$

where $e$ and $m$ are the electronic charge and mass respectively, and $a$ is the lengthscale associated with phase fluctuations which is typically of the order of the dirty limit coherence length, $\xi_0$. The central observation of this study is that in strongly disordered NbN films in the PG state, $J$ becomes dependent on the probing frequency and at high frequencies it continues to remain finite up to a temperature well above $T_c$. From comparison with STS measurements we identify this temperature with the PG temperature $T^*$ where the gap-like feature in the t-DOS disappears[11]. Our results support the scenario where a PG state consisting of phase incoherent domains is formed at $T^*$ and the superconducting state emerges from this PG state at $T_c$ when global phase coherence is established between all the domains.

Samples used in this study consist of a set of epitaxial NbN thin films with different levels of disorder grown on single crystalline MgO substrates using reactive dc magnetron sputtering. The disorder in the form of Nb vacancies in the NbN crystal lattice was controlled by controlling Nb/N ratio in the plasma. Details of sample preparation and characterization have been reported elsewhere[12]. For the samples used here, $T_c$, defined as the temperature at which dc resistance goes below our measurable limit varies in the range $T_c \approx 15.71 - 3.14$ K. The effective disorder, characterized using the product of the Fermi wave vector, $k_F$, and mean free path, $l$, are estimated from $T_c$ and normal state resistivity ($\rho$) to be in the range[11] $k_F l \sim 9.5 - 1.8$. The thickness ($t$) of all films determined using a stylus profilometer was ~50 nm. $\sigma(\omega)$ was measured using a home-built broadband Corbino microwave spectrometer operating the range 0.4 – 20 GHz, coupled to a continuous flow $^4$He cryostat operating down to 2.3 K. In this technique[10], concentric metallic contact pads of Au or Ag are deposited on the superconducting films in Corbino geometry such that the inner and outer electrodes on the sample mate with the inner and outer electrode of a modified microwave connector at the end of a transmission line (Fig. 1(a)). The sample thus acts as a terminator for an open ended transmission line. The complex reflection coefficient, $S_{11}$, is measured by measuring the reflected microwave signal from the sample using a vector network analyzer (Rhode and Schwartz). Using a standard procedure[13], a set of three samples with well known reflection coefficient were used to correct of extraneous reflections (at joints and bends in the transmission line), damping and phase shifts in the transmission line. A thick NbN film (300 nm) with $T_c$~16.1 K was used as a short standard ($S_{11} = -1$) at a temperature of 2.3K. A flat Teflon piece was used for open standard ($S_{11} = 1$). A NiCr evaporated film with 20Ω resistance between the two Corbino pads was used as a standard load. The sample impedance, $Z_s$, is obtained from the relation, $Z_s = Z_0 (1+S_{11}^c)/(1-S_{11}^c)$, where $S_{11}^c$ is the corrected reflection coefficient and $Z_0 = 50$ Ω is the characteristic impedance of our transmission line; $\sigma(\omega) = \frac{\ln(a/b)}{2\pi t Z_s}$ where $a$ and $b$ are the inner and outer diameters of the two "Corbino" electrodes. The resistivity ($\rho$) of the sample was measured in-situ using the bias-tee of the network analyzer using a two-probe technique. The temperature independent two-probe background resistance (~7 Ω) was subtracted while calculating the resistivity.

Figure 1(b)-(c) shows the representative data for $\sigma'(\omega)$ and $\sigma''(\omega)$ as a function of frequency at different temperatures for the sample with $T_c \sim$ 3.14 K. Consistent with the expected behavior in the superconducting state, at low temperatures $\sigma'(\omega)$ shows a sharp peak at $\omega \rightarrow 0$ whereas $\sigma''(\omega)$ varies as $1/\omega$ (dashed line). Well above $T_c$, $\sigma'(\omega)$ is flat and featureless and $\sigma''(\omega)$ is within the noise level of our measurement consistent with the behavior in a normal metal. Figure 2(a)-(d) shows $\sigma'(\omega)$-$T$, $\sigma''(\omega)$-$T$ and $J$-$T$ at different frequencies for four samples with different $T_c$. All samples display a dissipative peak in $\sigma'(\omega)$ close to $T_c$. For low disorder samples $\sigma''(\omega)$ for all frequencies dropped below our measurement limit close to $T_c$. As shown later the narrow fluctuation regime in these samples is well described by Aslamazov-Larkin (AL) theory of amplitude fluctuations. On the other hand samples with higher disorder show an extended fluctuation regime where $\sigma''(\omega)$ remains finite up to a temperature well above $T_c$. We convert $\sigma''(\omega)$ into $J$ (from eqn. 1) using the experimental values of $\xi_0$ obtained from upper critical field measurements[14]. For $T < T_c$, $J$ is frequency independent, showing that the phase rigidity is established over all time scales. However, for the samples with higher disorder (Fig. 2(c) and 2(d)), $J$ becomes strongly frequency dependent above $T_c$: While at 0.4 GHz $J$ falls to zero very close to $T_c$ with increase in frequency it acquires a long tail and remains finite well above $T_c$. To understand the connection between these observations and the PG state observed earlier in STS measurements[11], we compare $T^*$ with the temperature, $T_m^*$, at which $J$ goes below our measurable limit at 20 GHz. In Figure 3, we plot $T_m^*$ and $T_c$ for several samples along with the variation of $T^*$ and $T_c$ obtained from STS measurements, as a function of $k_F l$. Within the error limits of determining these temperatures, $T^* \approx T_m^*$, showing that the onset of the PG in the t-DOS and onset of finite $J$ at 20 GHz happen at the same temperature. Furthermore, only the samples in the disorder range where a PG state appears, show a difference between $T_c$ and $T_m^*$. We therefore attribute the frequency dependence of $J$ to a fundamental property related to the PG state.

Having established the relation between the PG state and the finite high frequency phase stiffness, we now look at the fluctuation region above $T_c$ more closely. On very general grounds, the fluctuating conductivity, $\sigma_{fl}(\omega)$ is predicted[15] to scale as, $\sigma_{fl}(\omega)/\sigma_{fl}(0) = S(\omega/\omega_0)$, where $\omega_0$ is the characteristic fluctuation frequency at that temperature. We obtain $\sigma_{fl}(\omega)$ from $\sigma(\omega)$ by subtracting the dc value of conductivity above $T_m^*$. Since the phase angle $\phi(\omega) = \tan^{-1}(\sigma''_{fl}(\omega)/\sigma'_{fl}(\omega))$ is the same as the phase angle of $S$, scaling $\phi(\omega)$ at each temperature with a different $\omega_0$ one can expect a collapse of all data on a single curve. For the amplitude the data would similarly scale when normalized by $\sigma_{fl}(0)$ for the same values of $\omega_0$ at each temperature. Such a scaling works for all the samples in the frequency range 0.4 – 12 GHz as shown in Figure 4(a) and 4(b) for the samples with $T_c$~15.71 K and 3.14 K respectively. We observe a very good consistency between the temperature variation of $\sigma_{fl}(0)$ and dc fluctuation conductivity, $\sigma_{fl}^{dc}$ obtained from $\rho$ - $T$ measured in the same run (Fig. 4(c) and 4(d)) showing the consistency of our scaling procedure. We can then have a closer look to the form of the scaling function $S$ obtained for our films. In general, the function $S$ is constrained by the physics of the low and high frequency limits: For $\omega \rightarrow 0$, $S \rightarrow 1$ corresponding to the normal state conductivity, and for $\omega \rightarrow \infty$, $S \rightarrow c(\omega/\omega_0)^{[(d-2)/z-1]}$ where c is a complex constant, d is the dimension and z is the dynamical exponent, with z=2 for models based on relaxational dynamics. This is the case for example of ordinary Ginzburg-Landau (GL) amplitude and phase fluctuations[16,17] which are between the possible candidates for the observed fluctuation conductivity, at least at moderate disorder[18]. Indeed, a direct comparison with the data for the sample with $T_c \sim 15.71$ K shows that $S$ matches very well with the Ashlamazov-Larkin (AL) prediction in d=2 dimensions, while Maki-Thompson (MT) corrections[19] are suppressed by disorder, in agreement with earlier measurements on low-disorder NbN films[18]. On the other hand, the corresponding curve for the sample with $T_c \sim 3.14$ K does not match with any of these models. Whereas for both samples $\omega_0 \to 0$ as $T \to T_c$, showing the critical slowing of fluctuations as the superconducting transition is approached, a clue as to the origin of this deviation is obtained from a comparison of the temperature variation of $\omega_0$ (Fig. 4(c) and 4(d)) with the prediction from 2D AL theory, i.e. $\omega_0^{AL} = \frac{16 k_B T_c}{\pi \hbar} \ln\left(\frac{T}{T_c}\right)$. Whereas for the film with $T_c \sim 15.71$ K the best scaling values of $\omega_0$ is in agreement with $\omega_0^{AL}$ within a factor of the order of unity, in the films with $T_c \sim 3.14$ K $\omega_0$ is more than one order of magnitude smaller than $\omega_0^{AL}$. The low characteristic fluctuation in the disordered sample signals a fundamental breakdown of the amplitude fluctuation scenario, which cannot be accounted for by any simple adjustment of parameters in these models.

We can now put these observations in perspective. It has been shown from STS measurements that in the presence of strong disorder the spatial landscape of the superconductor become highly inhomogeneous, thereby forming domain like structures, tens of nm in size, where the superconducting OP is finite and regions where the OP is completely suppressed[7]. One way to visualize the superconducting state is as a disordered network of Josephson junctions, where the domains observed in STS get coupled through Josephson coupling giving rise to the global phase coherent state. In this scenario, $T_c$ corresponds to the temperature at which the weakest couplings are broken. Therefore, just above $T_c$ the sample consist of large phase coherent domains (consisting of several smaller domains) fluctuating with respect to each other. We believe the large fluctuations at low frequency in our most disordered sample originate from fluctuations between coherent domains which gradually fragment with increasing temperature. As the temperature is increased further, the large domains progressively fragment eventually reaching the limiting size observed in STS measurements at a temperature close to $T^*$. In such a scenario $J$ will depend on the length scale at which it is probed. When probed on a length scale much larger than the phase coherent domains, $J \rightarrow 0$. On the other hand, when probed at length scale of the order of the domain size $J$ would be finite. $J(\omega)$ probes the phase stiffness over a length scale set by the diffusion of the electron over one cycle of radiation which is given by $L(\omega) = [D/(\omega/2\pi)]^{0.5}$. Here, $D$, is the electronic diffusion constant given by, $D \approx v_F l/d$, where $v_F$ and $l$ are the Fermi velocity and the electronic mean free path[20] and $d$ is the dimension. Therefore, for a given frequency, $J(\omega)$ would vanish at a temperature where the phase coherent domain becomes much smaller than $L(\omega)$. Using this criterion for each temperature, we define the characteristic length scale of the phase coherent domains, $L_0(T) = L(\omega)$, corresponding to the frequency for which $J(\omega)$ goes below measurement resolution (Fig 2 (c) and 2(d)). Taking $d=3$ (since $l \ll t$), the limiting value of $L_0$ at $T \approx T^*$ is between 50-60 nm which is in the same order of magnitude as the domains observed in STS measurements[7] on NbN films with similar $T_c$.

In summary, we have shown that in strongly disordered NbN thin films which display a PG state, $J$ becomes dependent on the temporal and spatial length scale for $T_c < T < T^*$. The remarkable consistency between $T^*$ determined from STS and microwave measurements, and the evident deviations from the expected behavior for GL fluctuations, suggest the notion that the superconducting transition in these systems is driven by phase disordering of a strongly inhomogeneous superconducting background. While our results bear strong similarity with the results obtained earlier on a disordered $InO_x$ film[21], we would like to note that the whole fluctuation regime cannot be attributed to Kosterlitz-Thouless (KT) physics, since earlier measurements[22] have shown that the KT fluctuation regime is restricted to a narrow temperature range above $T_c$. On the other hand, the persistence far from $T_c$ of slow relaxation processes, as characterized by the low frequency scale $\omega_0$, resembles the emergence of a glassy physics, which has been recently predicted at strong disorder[23,24] and indirectly observed by STS[5,7]. However, an explicit connection between our findings and existing predictions, as for example the relevance of a small number of percolative paths for the current at strong disorder[24], is still lacking, and its formulation would certainly improve considerably our understanding of the superconductor-insulator transition.

*Acknowledgement:* We thank Peter Armitage for valuable suggestions regarding the experiment and analysis, and Claudio Castellani and Andrey Varlamov for useful discussions.

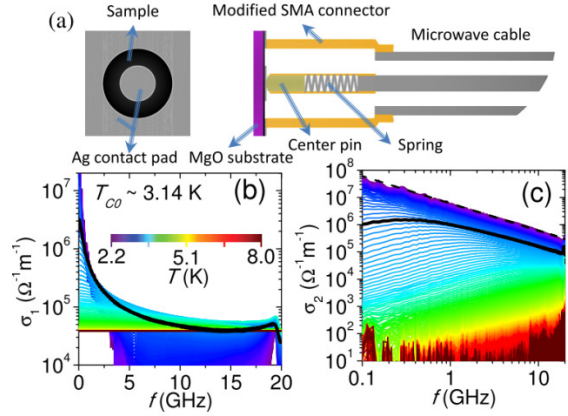

**Figure 1.** Schematic drawing of the superconducting film with Ag contact pads in Corbino geometry (left) and of the 50 Ω transmission line terminating with the sample (right); a specially modified microwave connector with spring loaded centre pin is used to establish the contact with the inner and outer contact pads. (b) $\sigma'(\omega)$ and (c) $\sigma''(\omega)$ at different temperatures as a function of frequency; the solid black line shows the data at $T = T_c$. The color scale corresponding to the temperatures is shown in panel (b). The dashed line in (c) is a fit to $\sigma''(\omega) \propto 1/\omega$ for the data at 2.24 K.

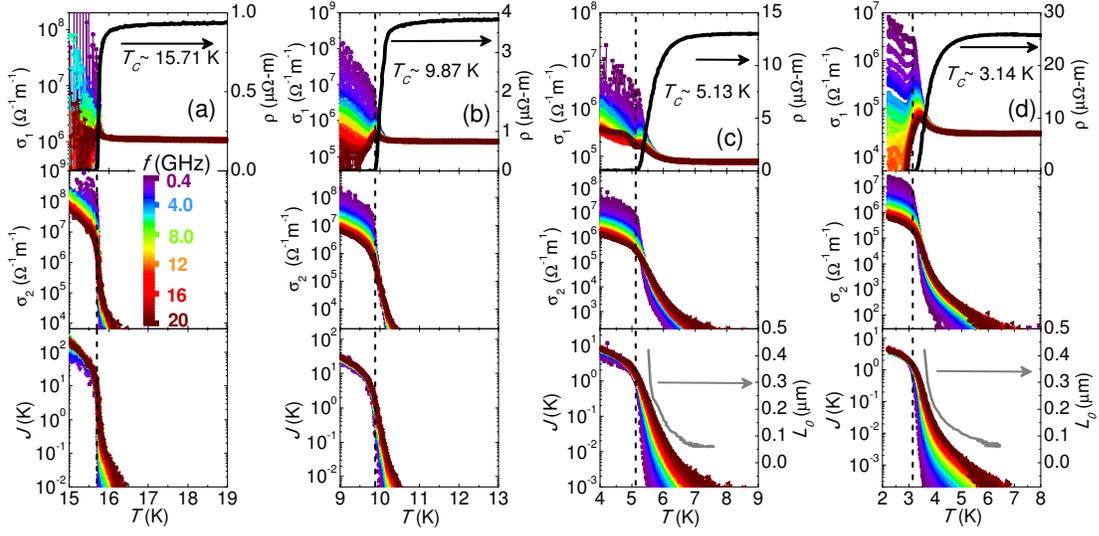

**Figure 2.** Temperature dependence σ' (upper panel), σ" (middle panel) and $J$ (lower panel) at different frequencies for four samples with (a) $T_c \sim 15.7$ K (b) $T_c \sim 9.87$ K (c) $T_c \sim 5.13$ K and (d) $T_c \sim 3.14$ K. The color scale displaying different frequencies is displayed in (a). The solid (black) lines in the top panels show the temperature variation of resistivity. Vertical dashed lines correspond to $T_c$. The solid (gray) lines in the bottom panels of (c) and (d) show the variation of $L_0$ above $T_c$.

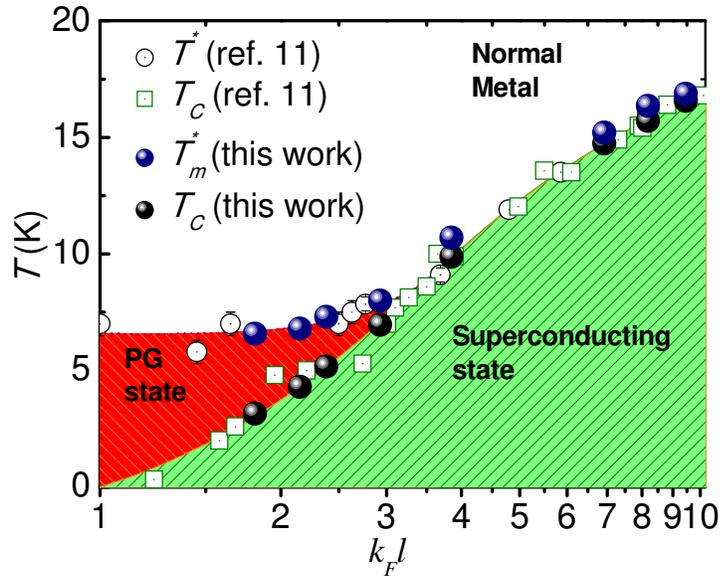

**Figure 3.** Phase diagram showing $T_c$ and $T^*$ obtained from STS measurements (Ref. 11) along with $T_c$ and $T_m^*$ obtained from microwave measurements.

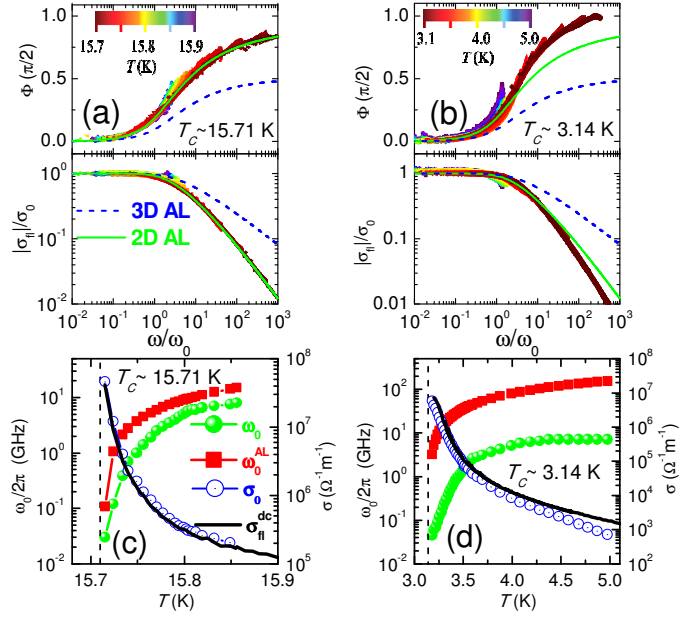

**Figure 4.** (a)-(b) Rescaled phase (upper panel) and amplitude (lower panel) of $\sigma_{fl}(\omega)$ using the dynamic scaling analysis on two films with $T_c \sim 15.7$ and $3.14$ K respectively. The solid lines show the predictions from 2D and 3D AL theory. The color coded temperature scale for the scaled curves is shown in each panel. (c)-(d) Variation of $\omega_0$, $\omega_0^{AL}$, $\sigma_0$ and $\sigma_{dc}$ as a function of temperature. The dashed vertical lines correspond to $T_c$.

# Supplementary Material

## I. Ashlamazov-Larkin (AL) and Maki-Thompson (MT) expressions for fluctuation conductivity

The contribution of AL and MT fluctuations to the dc conductivity in 2D and 3D are given by[1,2]

$$\sigma_{dc}^{2D-AL} = \frac{1}{16} \frac{e^2}{\hbar t} \varepsilon^{-1} \tag{1}$$

$$\sigma_{dc}^{3D-AL} = \frac{1}{32} \frac{e^2}{\hbar \xi_0} \varepsilon^{-1/2} \tag{2}$$

$$\sigma_{dc}^{2D-MT} = \frac{1}{8} \frac{e^2}{\hbar t} \frac{1}{\varepsilon - \delta} \ln\left(\frac{\varepsilon}{\delta}\right) \tag{3}$$

$$\sigma_{dc}^{3D-MT} = \frac{1}{8} \frac{e^2}{\hbar \xi_0} \varepsilon^{-1/2} \tag{4}$$

where $\varepsilon = \ln(T/T_c)$, $t$ is the thickness of the sample and $\xi_0$ is the BCS coherence length and $\delta$ is the Maki-Thompson pair breaking parameter. The two contributions are additive.

The frequency dependence of the fluctuation conductivities are as follows.

*Ashlamazov-Larkin[3]:*

$$\sigma^{2DAL}(\omega) = \sigma_{DC}^{2DAL} S^{2DAL}\left(\frac{\omega}{\omega_0}\right); \omega_0 = \frac{16 k_B T_c}{\pi \hbar} \varepsilon$$

$$S^{2DAL}(x) = \left\{\frac{2}{x} \tan^{-1} x - \frac{1}{x^2} \ln(1+x^2)\right\} + i\left\{\frac{2}{x^2}(\tan^{-1} x - x) + \frac{1}{x} \ln(1+x^2)\right\} \tag{5}$$

$$\sigma^{3DAL}(\omega) = \sigma_{DC}^{3DAL} S^{3DAL}\left(\frac{\pi \hbar \omega}{16 k_B T_c \varepsilon}\right)$$

$$S^{3DAL}(x) = \left\{\frac{8}{3x^2}\left(1 - (1+x^2)^{3/4} \cos(\frac{3}{2} \tan^{-1} x)\right)\right\} + i\left\{\frac{8}{3x^2}\left(-\frac{3}{2}x + (1+x^2)^{3/4} \sin(\frac{3}{2} \tan^{-1} x)\right)\right\} \tag{6}$$

*Ashlamazov-Larkin and Maki-Thompson[4]:*

$$\sigma^{2DAL+MT}(\omega) = \sigma_{DC}^{2DAL} S^{2DAL+MT}\left(\frac{\pi\hbar\omega}{16k_BT_c\varepsilon}\right)$$

$$S^{2DAL+MT}(x) = \left\{\text{Re}\, S^{2DAL}(x) + \frac{2\pi x - 2\ln(2x)}{1+4x^2}\right\} + i\left\{\text{Im}\, S^{2DAL}(x) + \frac{\pi + 4x\ln(2x)}{1+4x^2}\right\}$$

(7)

$$\sigma^{3DAL+MT}(\omega) = \sigma_{DC}^{3DAL} S^{3DAL+MT}\left(\frac{\pi\hbar\omega}{16k_BT_c\varepsilon}\right)$$

$$S^{3DAL+MT}(x) = \left\{\text{Re}\, S^{3DAL}(x) + \frac{4 - 4x^{1/2} + 8x^{3/2}}{1+4x^{1/2}}\right\} + i\left\{\text{Im}\, S^{3DAL}(x) + \frac{4x^{1/2} - 8x + 8x^{3/2}}{1+4x^{1/2}}\right\}$$

(8)

(We follow the same notation as in ref. 5.)

In Figure 1s, we show below the scaled phase and amplitude for the sample with 15.71 K along with the predicted theoretical variation from eq. (5)-(8).

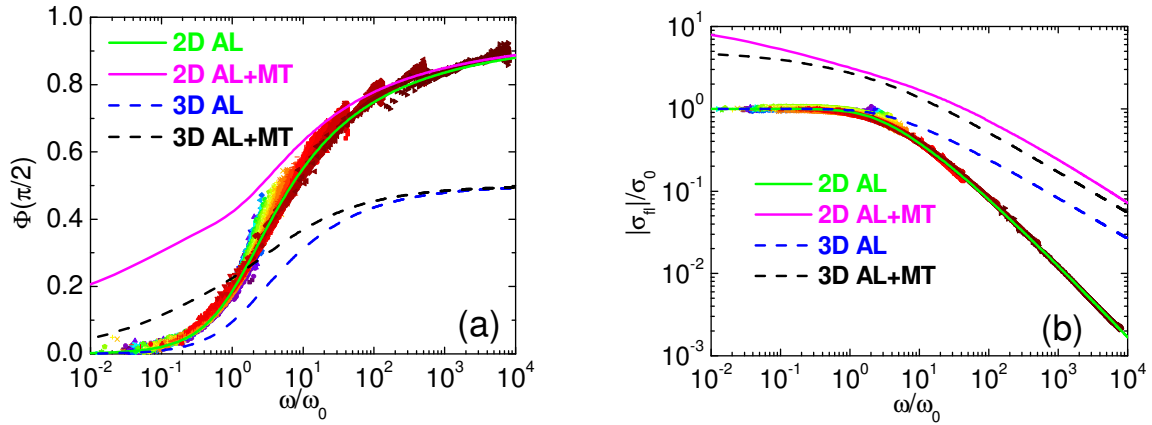

**Figure 1s. Scaled (a) phase and (b) amplitude of the fluctuation conductivity for NbN thin film with $T_c \sim 15.71$ K along with theoretical predictions for AL and AL+MT theory in 2D and 3D.**

## II. Fermi velocity ($v_F$), electronic mean free path ($l$) *and* diffusion constant ($D$) for NbN thin films with different $T_c$

We have determined the values $v_F$ and $l$ for a large number of epitaxial NbN films with different levels of disorder grown on single crystalline MgO substrates, similar to the ones used in the present study. $v_F$ and $l$ (and $k_Fl$) were determined from the resistivity and Hall coefficient[6] measured at 285K using free-electron formulae. These values are listed in Table 1 along with the electronic diffusion constant, $D$, and the characteristic probing length-scales, $L(\omega)$, at 0.4 GHz and 20 GHz.

**Table 1.**

| $k_F l$ | $T_c$ (K) | $v_F$ (m s$^{-1}$) | $l$ (Å) | $D_{diff}$ (m$^2$ s$^{-1}$) | $L(\omega)$ (nm) for 0.4 GHz | $L(\omega)$ (nm) for 20 GHz |
|---|---|---|---|---|---|---|
| 10.17 | 16.5 | 2260000 | 5.18 | 3.90E-04 | 987.7 | 139.7 |
| 8.7 | 16.1 | 2150000 | 4.69 | 3.36E-04 | 916.6 | 129.6 |
| 7.09 | 15 | 2030000 | 4.04 | 2.73E-04 | 826.7 | 116.9 |
| 5.5 | 13.5 | 1910000 | 3.34 | 2.13E-04 | 729.1 | 103.1 |
| 4.6 | 11.6 | 1830000 | 2.94 | 1.79E-04 | 669.6 | 94.7 |
| 4.38 | 11.1 | 1800000 | 2.83 | 1.70E-04 | 651.5 | 92.1 |
| 3.91 | 10 | 1750000 | 2.62 | 1.53E-04 | 618.1 | 87.4 |
| 3.49 | 9.02 | 1690000 | 2.42 | 1.36E-04 | 583.8 | 82.6 |
| 2.85 | 6.99 | 1610000 | 2.09 | 1.12E-04 | 529.5 | 74.9 |
| 2.21 | 4.78 | 1510000 | 1.72 | 8.66E-05 | 465.2 | 65.8 |
| 2 | 3.97 | 1470000 | 1.59 | 7.79E-05 | 441.3 | 62.4 |
| 1.23 | 0.6 | 1340000 | 1.06 | 4.73E-05 | 344 | 48.7 |